\documentclass[11pt]{article}
\usepackage{moriond,epsfig}

\def\Journal#1#2#3#4{{#1} {\bf #2}, #3 (#4)}

\def\EPJC{{\em Eur. Phys. J.} C}
\def\NIMA{{\em Nucl. Instrum. Methods} A}
\def\PLB{{\em Phys. Lett.}  B}
\def\PRL{\em Phys. Rev. Lett.}
\def\PRD{{\em Phys. Rev.} D}
\def\PTP{\em Prog. Theor. Phys.}
\def\JPG{{\em J. Phys.} G}
\def\JINST{\em JINST}

\RequirePackage{xspace}
\def\cbar   {\ensuremath{\overline c}\xspace}
\def\KS     {\ensuremath{K^0_{\scriptscriptstyle S}}\xspace} 
\def\KL     {\ensuremath{K^0_{\scriptscriptstyle L}}\xspace} 
\def\Dbar   {\kern 0.2em\overline{\kern -0.2em D}{}\xspace}
\def\Bbar   {\kern 0.18em\overline{\kern -0.18em B}{}\xspace}
\def\BB     {\ensuremath{B\Bbar}\xspace} 
\def\jpsi   {\ensuremath{{J\mskip -3mu/\mskip -2mu\psi\mskip 2mu}}\xspace}
\def\psitwos{\ensuremath{\psi{(2S)}}\xspace}
\mathchardef\Upsilon="7107
\def\Y#1S{\ensuremath{\Upsilon{(#1S)}}\xspace}

\def\BF{{\ensuremath{\cal B}}\xspace}
\newcommand{\stat}{\ensuremath{\rm{(stat)}}\xspace}
\newcommand{\syst}{\ensuremath{\rm{(syst)}}\xspace}
\def\CP {\ensuremath{C\!P}\xspace}
\newcommand{\figref}[1]{Figure~\ref{fig:#1}}

\newcommand{\eqref}[1]{Eq.~\ref{eq:#1}}
\def\invfb{\ensuremath{\mbox{\,fb}^{-1}}\xspace}

\begin{document}
\vspace*{4cm}
\title{Recent {\boldmath $\CP$} violation results from Belle}

\author{Gagan B. Mohanty}

\address{Tata Institute of Fundamental Research, Homi Bhabha Road,\\
Colaba, Mumbai 400 005, India}

\maketitle\abstracts{
We summarize recent results on an array of $\CP$ violation measurements
performed by the Belle experiment using the data collected near the
$\Y4S$ and $\Y5S$ resonances at the KEKB asymmetric-energy $e^+e^-$
collider.}

\section{Introduction}

Flavor sector is the largest contributor in terms of number of free
parameters to the standard model (SM) of elementary particles -- a
confluence of electroweak interactions and quantum chromodynamics. In
particular, the phenomenon of quark-flavor mixing is described by three Euler
angles and one irreducible phase of the so-called Cabibbo-Kobayashi-Maskawa
(CKM) matrix~\cite{ckm}. This phase is the lone source of $\CP$ violation
within the SM. Unitarity of the $3\times 3$ CKM matrix leads to a set of
relations among its different elements that can be represented as triangles
in the complex plane. One such triangle, better known as the unitarity
triangle (UT), is the pictorial sketch of the condition $V_{ud}V^*_{ub}+
V_{cd}V^*_{cb}+V_{td}V^*_{tb}=0$. The raison d'$\hat{e}$tre of the two
$B$-factory experiments, Belle~\cite{belle} at KEK and BaBar~\cite{babar}
at SLAC, has been to precisely measure the sides and angles ($\phi_1$,
$\phi_2$ and $\phi_3$) \footnote{Another notation of $\beta$, $\alpha$ and
$\gamma$ is also available in the literature.} of the UT. The underlying
idea is to check the overall consistency of the CKM framework; any
significant discrepancy between various measurements could be interpreted
as potential new physics effects. In these proceedings, we report on
recent $\CP$ violation measurements carried out with Belle using $e^+e^-$
collision data collected near the $\Y4S$ and $\Y5S$ resonances.

\section{Detector and data}

Belle is a large-solid-angle magnetic spectrometer that operated at the
KEKB asymmetric-energy $e^+e^-$ collider. Before stopping the operation
in June 2010 to make way for its approved upgrade, Belle II~\cite{belle2},
the experiment has succeeded in accumulating a large data sample spread
over various bottomonium resonances -- it in fact holds the current world
record for the $\Y2S$, $\Y4S$ and $\Y5S$ samples. Unless stated otherwise,
results presented here comprise the full $\Y4S$ [$711\invfb$ corresponding
to $772\times 10^6\BB$ pairs] and $\Y5S$ [$121\invfb$] data collected with
Belle.

\section{Methodology}

The UT angles $\phi_1$ and $\phi_2$ are determined through the measurement
of time-dependent $\CP$ asymmetry,
\begin{equation}
A_{\CP}(\Delta t)=\frac{N[\Bbar^0(\Delta t)\to f_{\CP}]-N[B^0(\Delta t)\to f_{\CP}]}
                       {N[\Bbar^0(\Delta t)\to f_{\CP}]+N[B^0(\Delta t)\to f_{\CP}]},
\label{eq:acp1}
\end{equation}
where $N[\Bbar^0/B^0(\Delta t)\to f_{\CP}]$ is the number of $\Bbar^0/B^0$s
that decay into a $\CP$ eigenstate $f_{\CP}$ after a proper-time interval
$\Delta t$, starting with the decay of the other $B$ in the event
($B_{\rm tag}$). The $B_{\rm tag}$ daughters identify its flavor at the
decay time. The asymmetry, in general, can be expressed as a sum of two
terms:
\begin{equation}
A_{\CP}(\Delta t)=S_f\sin(\Delta m\Delta t)+A_f\cos(\Delta m\Delta t),
\label{eq:acp2}
\end{equation}
where $\Delta m$ is the $B^0\Bbar^0$ mixing frequency. The sine coefficient
$S_f$ is related to the UT angles, while the cosine coefficient $A_f$ is a
measure of direct $\CP$ violation. For the $A_f$ to have a nonzero value,
we need at least two competing amplitudes with different weak and strong
phases to contribute to the decay final state. The measurement of the UT
angle $\phi_3$ mostly inherits from the nonzero direct $\CP$ violation in
some charged $B$ meson decays.

\section{{\boldmath $\sin(2\phi_1)$} in {\boldmath $B^0\to (c\cbar)K^0$} decays}
\label{sec:phi1}

The most precise determination of the angle $\phi_1$ is provided by the
$B^0\to (c\cbar)K^0$ decays. These decays, known as ``golden modes'',
mainly proceed via the Cabibbo-favored tree diagram $b\to c\bar{c}s$
with an internal $W$ boson emission. The subleading penguin (loop)
contribution to the final state, which has a different weak phase
compared to the tree-level transition, is suppressed by almost two
orders of magnitude. This makes $A_f=0$ in \eqref{acp2} to a very
good approximation. Besides very small theoretical uncertainty
involved, these channels also offer experimental advantages because
of the relatively large branching fractions ($\sim 10^{-3}$) and the
presence of narrow resonances in the final state, which provides a
powerful discrimination against the combinatorial background.

\begin{figure}
\begin{center}
\includegraphics[width=0.48\columnwidth]{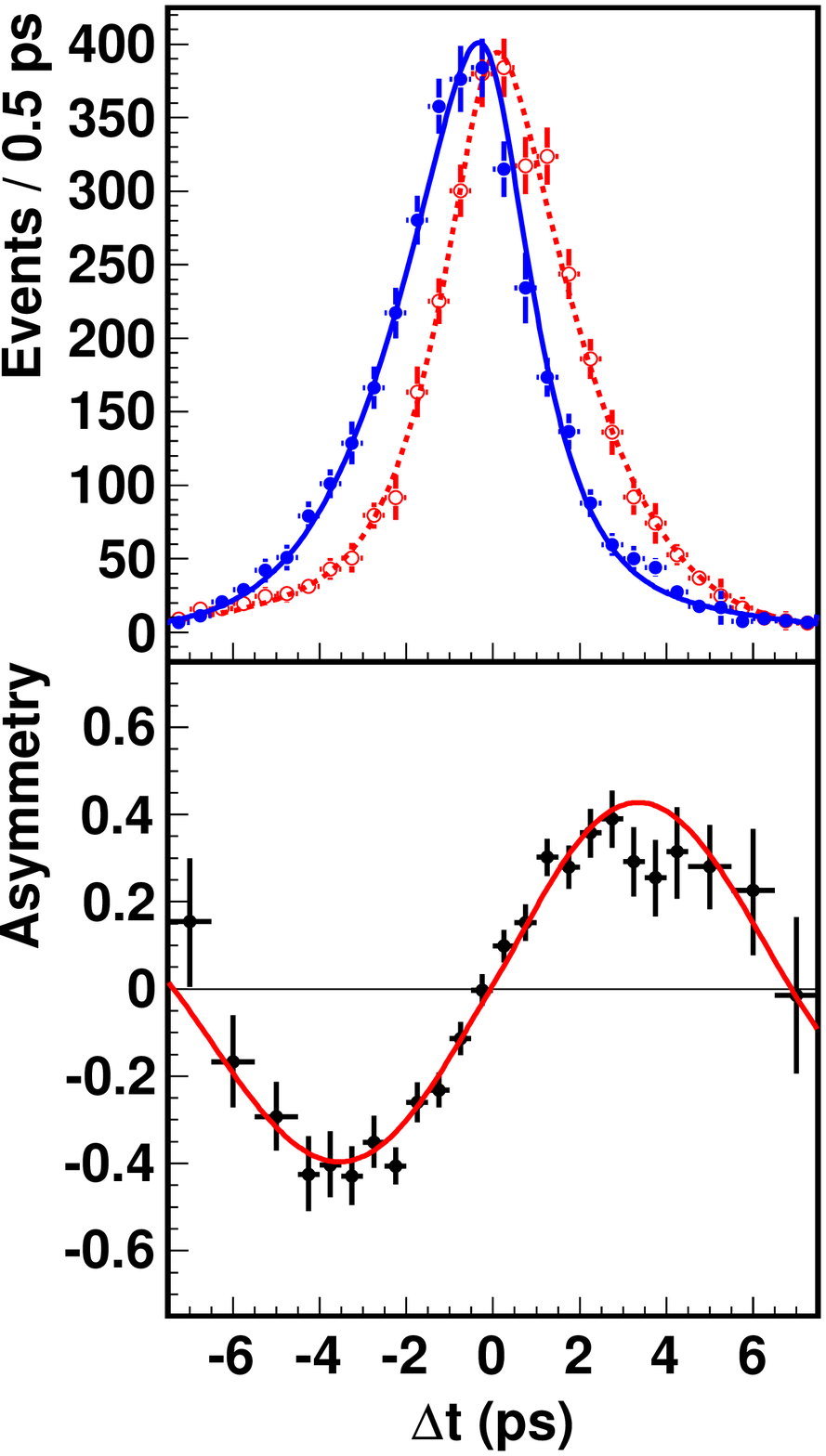}
\includegraphics[width=0.48\columnwidth]{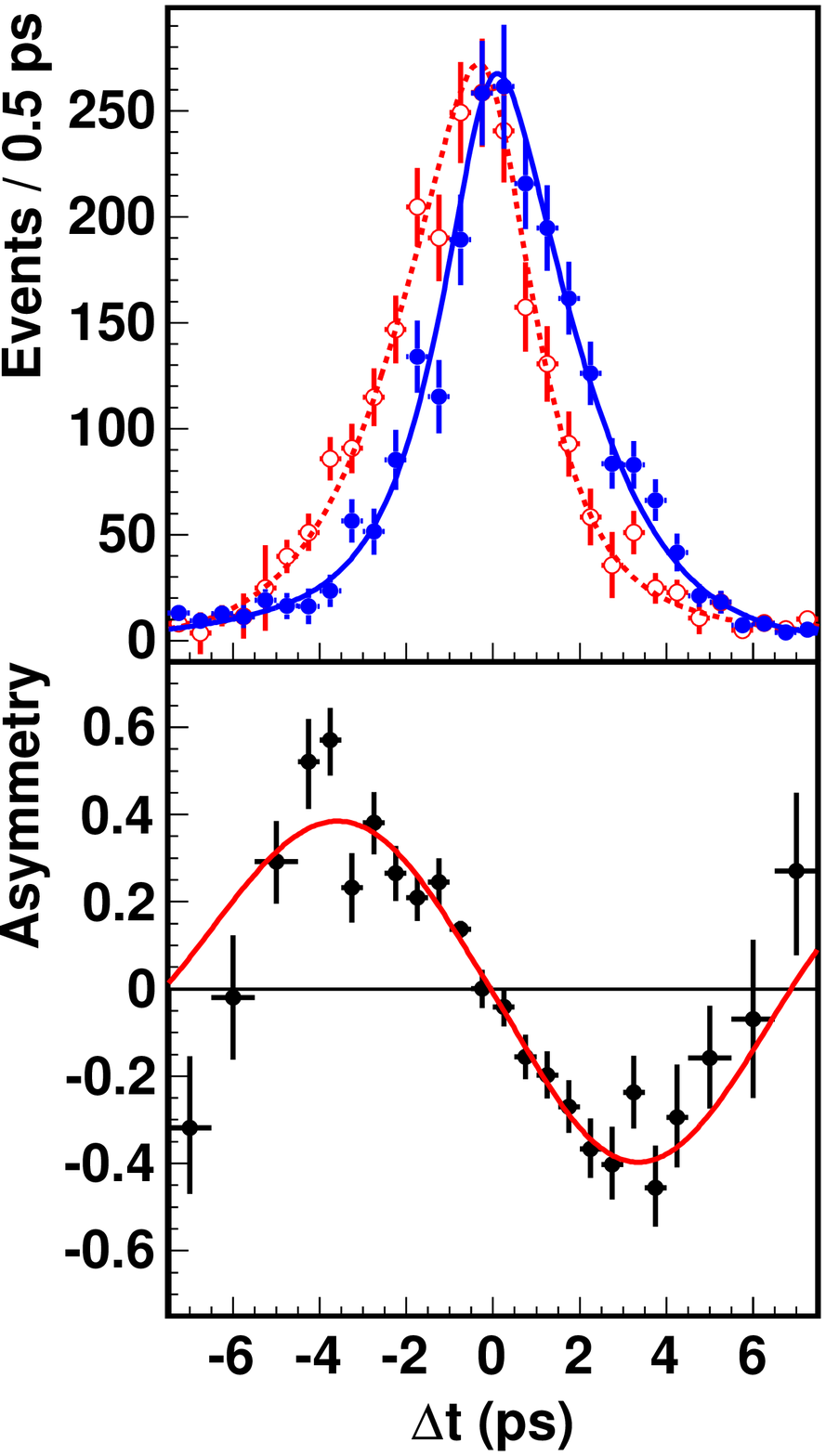}
\end{center}
\caption{Background-subtracted proper-time distributions (top) for
$\Bbar^0$ (blue solid) and $B^0$ (red dotted) tagged events and $\CP$
asymmetry (bottom) for good tag quality events for all considered
$\CP$-odd modes combined (left) and the $\CP$-even mode (right).}
\label{fig:sin2phi1}
\end{figure}

Belle has updated its previous results~\cite{oldphi1} on $\sin(2\phi_1)$
with the entire $\Y4S$ data set. In addition to more data, an improved
track reconstruction algorithm has yielded significant enhancement
in the reconstruction efficiency, e.g., about $18\%$ improvement in
the $B^0\to\jpsi\KS$ channel. The $\CP$ eigenstates considered in the
analysis are $\jpsi\KS$, $\psitwos\KS$, $\chi_{c0}\KS$ (all $\CP$
odd), and $\jpsi\KL$ ($\CP$ even). \figref{sin2phi1} shows $\Delta t$
distributions and time-dependent $\CP$ asymmetries for the candidate
events that satisfy good tag quality. The observed asymmetry
pattern is consistent with the $\CP$ eigenvalue, and there is a
negligible height difference -- a measure of direct $\CP$ violation
-- between $B^0$ and $\Bbar^0$ decays. We measure $\sin(2\phi_1)=
0.667\pm0.023\stat\pm0.012\syst$ and $A_f=0.006\pm0.016\stat\pm0.012
\syst$~\cite{newphi1}. This constitutes the most precise determination
of mixing-induced $\CP$ violation in a $B$ meson decay, and hence
provides a solid reference point for the SM that can be used to
test for evidence of new physics. 

\section{{\boldmath $\sin(2\phi_1)$} with the {\boldmath $\Y5S$} data}

Belle has measured the $\CP$-violation parameter $\sin(2\phi_1)$ using
a new method called the ``$B$-$\pi$ tagging'', where the charge of the
pion in $\Y5S\to\Bbar^{(*)0}B^{(*)+}\pi^-$ decays determines the
flavor of the accompanying neutral $B$ meson. The neutral $B$ is
fully reconstructed in some $\CP$-specific final state, and there is
no need to explicitly reconstruct the charged $B$ meson. Rather, it
can be indirectly inferred through the recoil (missing) mass of the
neutral $B$ and the pion, thanks to the precise knowledge of the energy
and momentum of initial states in an $e^+e^-$ $B$-factory. One can then
extract $\sin(2\phi_1)$ from the time-integrated asymmetry of the
$\pi^+$ and $\pi^-$ tagged events:
\begin{equation}
A_{BB\pi}=\frac{N_{BB\pi^-}-N_{BB\pi^+}}{N_{BB\pi^-}+N_{BB\pi^+}}=\frac{Sx+A}{1+x^2}
\label{eq:Bpi}
\end{equation}
where $S$ $(A)$ is the mixing-induced (direct) $\CP$ violation parameter,
$x=(m_H-m_L)/\Gamma$ is the mixing parameter with $m_H$ $(m_L)$ is the
mass of the heavy (light) neutral $B$ mass eigenstate and $\Gamma$ is their
average decay width.

\begin{figure}
\begin{center}
\includegraphics[width=0.48\columnwidth]{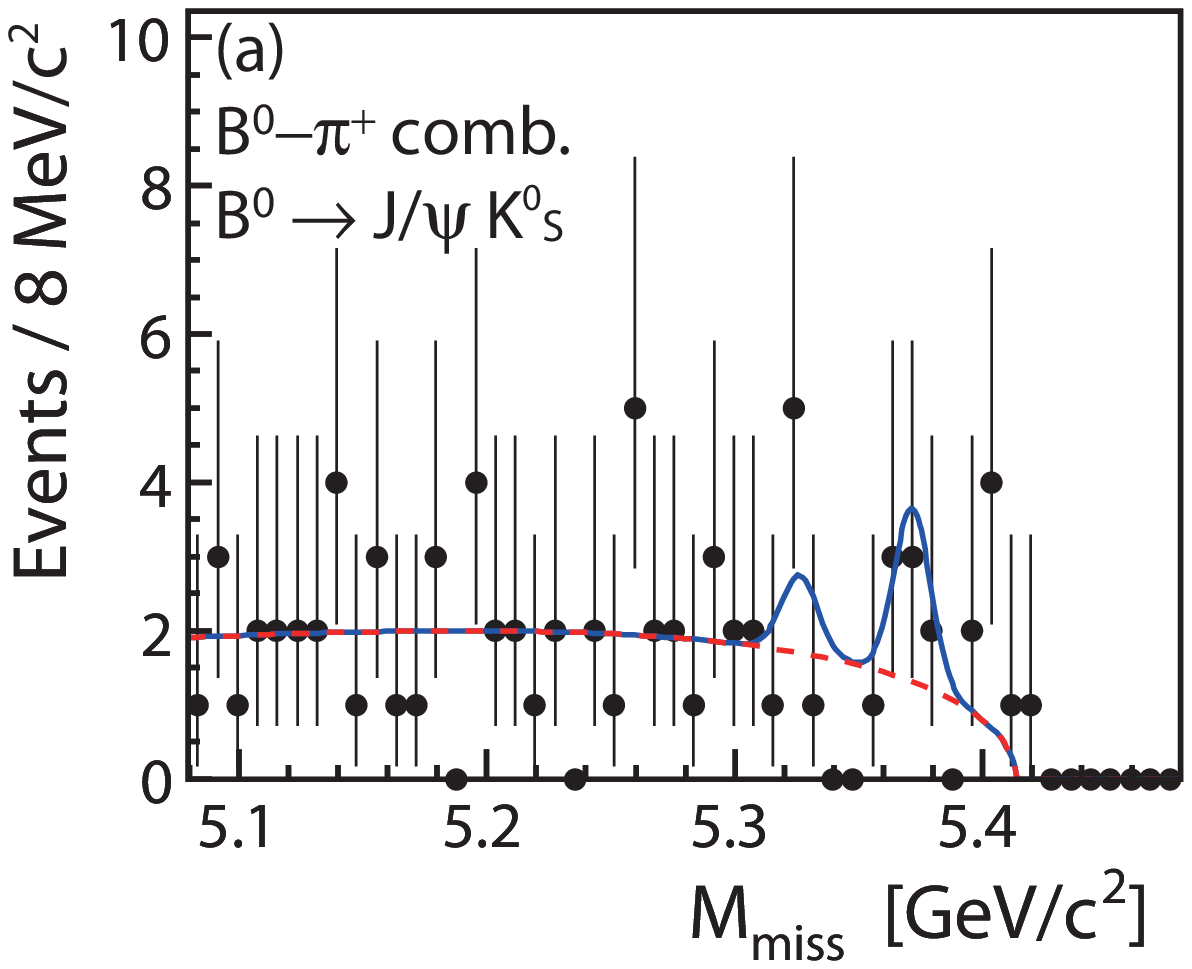}
\includegraphics[width=0.48\columnwidth]{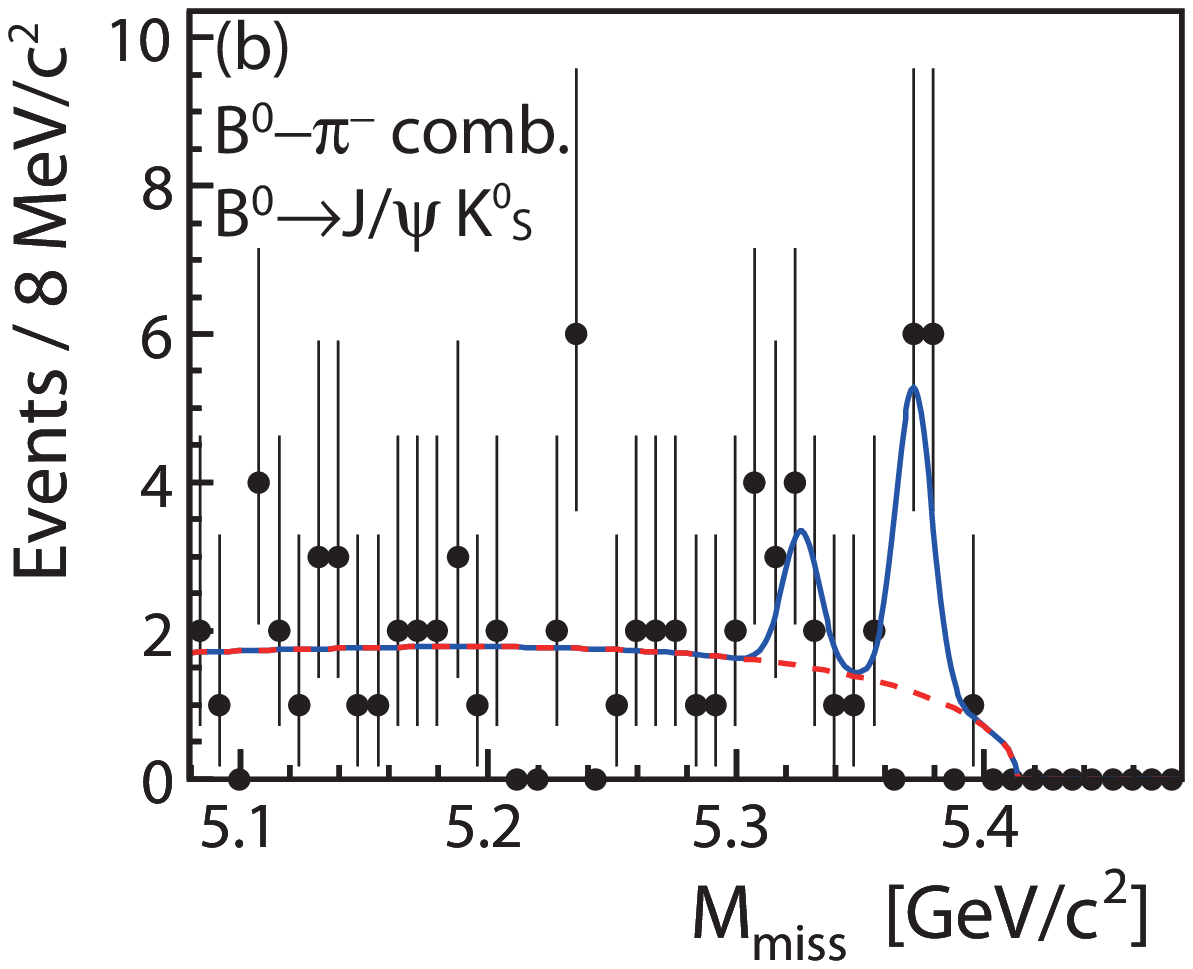}
\end{center}
\caption{Missing mass distributions for $B^0\to\jpsi\KS$ candidates
tagged by (a) $\pi^+$ and (b) $\pi^-$ in the $\Y5S$ data sample. Points
with error bars are the data, solid curves are the fit projections for
signal-plus-background, and dashed curves show the background contribution.}
\label{fig:sin2phi1y5s}
\end{figure}

\figref{sin2phi1y5s} shows the missing mass distributions of
neutral $B$ mesons reconstructed in the channel $B^0\to\jpsi\KS$
together with the charged pion, separately for the $\pi^+$ and $\pi^-$
tagged samples. Two peaks denote $B\Bbar^*\pi+B^*\Bbar\pi$ (first) and
$B^*\Bbar^*\pi$ (second) contributions. Expectedly, they are separated
by the mass difference between a $B^*$ and a $B$ meson. The fit to the
available $\Y5S$ sample yields the number of $\pi^+$ ($\pi^-$) tagged
events to be $7.8\pm3.9$ ($13.7\pm5.3$); from this we determine
$A_{BB\pi}=0.28\pm0.28\stat$. Assuming direct $\CP$ violation term
$A$ to be zero (consistent with the $B^0\to (c\cbar)K^0$ results) and
using the world-average value~\cite{pdg} of the mixing parameter $x$
$(0.771\pm0.007)$ in \eqref{Bpi}, we obtain $\sin(2\phi_1)=0.57\pm
0.58\stat\pm0.06\syst$~\cite{phi1y5s}.

The $B$-$\pi$ tagging method is complimentary to the time-dependent
analyses carried out at the $\Y4S$ peak with the flavor tagging of
neutral $B$ mesons. Although at the moment we are limited by the
available $\Y5S$ statistics, this has a great potential for the
super flavor factory experiments such as Belle II.

\section{Measurement of the angle {\boldmath $\phi_3$}}
\label{sec:phi3}

The UT angle $\phi_3$ unlike $\phi_1$ (discussed in Section~\ref{sec:phi1})
relies on the measurement of direct $\CP$ violation in $B^-\to D^{(*)}K^-$
decays caused by interference between the two contributing amplitudes,
where both $D^0$ and $\Dbar^0$ mesons decay to a common final state.
The fact that one of the amplitudes is almost an order of magnitude
smaller than the other ($B^-\to\Dbar^0K^-$ and $B^-\to D^0K^-$,
respectively) make our life difficult on extracting $\phi_3$. The
measurements are performed in three different ways: (a) by utilizing
decays of $D$ mesons to $\CP$ eigenstates, such as $\pi^+\pi^-$,
$K^+K^-$ ($\CP$ even) or $\KS\pi^0$, $\phi\KS$ ($\CP$ odd)~\cite{glw},
(b) by making use of doubly Cabibbo suppressed decays of $D$ mesons,
e.g., $D^0\to K^+\pi^-$~\cite{ads}, or (c) by exploiting the interference
pattern in the Dalitz plot (DP) of the $D$ decays such as $D\to\KS\pi^+
\pi^-$~\cite{ggsz}. The first two methods are theoretically clean but
suffer from low statistics. The Dalitz method at present provides the
strongest constraint on $\phi_3$. In the following two subsections,
we describe recent updates on $\phi_3$ from Belle.

\subsection{Model-independent Dalitz plot analysis}

Using a model-dependent DP method, Belle's earlier measurement~\cite{dalitz}
based on a data sample of $605\invfb$ integrated luminosity yielded $\phi_3=
(78.4^{+10.8}_{-11.6}\pm3.6\pm8.9)^\circ$ and $r_B=0.160^{+0.040}_{-0.038}
\pm0.011^{+0.050}_{-0.010}$, where the uncertainties are statistical,
experimental systematic and DP model dependence, respectively. ($r_B$
is the ratio of the amplitudes for $B^-\to\Dbar^0K^-$ and $B^-\to D^0
K^-$.) Although with more data one can squeeze on the statistical part,
the result will still remain limited by the model uncertainty.

In a bid to circumvent this problem, Belle has carried out a
model-independent analysis, following the idea originally proposed
by Giri {\it et al.}~\cite{ggsz} that is further extended in a
latter work~\cite{bonder}. The analysis is based on the full $\Y4S$
data sample. In contrast to the conventional DP method, where the
$D\to\KS\pi^+\pi^-$ amplitudes are parameterized as a coherent
sum of several quasi-two-body amplitudes as well as a nonresonant
term, the model-independent approach invokes study of a binned DP.
In this approach, the expected number of events in the $i^{\rm th}$
bin of the DP for the $D$ mesons from $B^{\pm}\to DK^{\pm}$ is given
by
\begin{equation}
N^{\pm}_i=h_B[K_{\pm i}+r^2_BK_{\mp i}+2\sqrt{K_iK_{-i}}
(x_{\pm}c_i\pm y_{\pm}s_i)],
\label{eq:midp}
\end{equation}
where $h_B$ is the overall normalization and $K_i$ is the number
of events in the $i^{\rm th}$ DP bin of the flavor-tagged (whether
$D^0$ or $\Dbar^0$) $D\to\KS\pi^+\pi^-$ decays, accessible via the
charge of the slow pion in $D^{*\pm}\to D\pi^{\pm}$. The terms
$c_i=<\!\!\cos\Delta\delta_D\!\!>$ and $s_i=
<\!\!\sin\Delta\delta_D\!\!>$ contain information about the
strong-phase difference between the symmetric DP points
[$m^2(\KS\pi^+),m^2(\KS\pi^-)$] and [$m^2(\KS\pi^-),m^2(\KS\pi^+)$];
they are the external inputs obtained from quantum correlated
$D^0\Dbar^0$ decays at the $\psi(3770)$ resonance in
CLEO~\cite{libby}. Finally $x_{\pm}=r_B\cos(\delta_B\pm\phi_3)$
and $y_{\pm}=r_B\sin(\delta_B\pm\phi_3)$, where $\delta_B$ is the
strong-phase difference $B^- \to\Dbar^0K^-$ and $B^-\to D^0K^-$.

\begin{table}[t]
\caption{Results on the $x,y$ parameters and their statistical
correlation for $B\to DK$ decays. The quoted uncertainties are
statistical, systematic and precision on $c_i,s_i$,
respectively.\label{tab:midp}}
\vspace{0.4cm}
\begin{center}
\begin{tabular}{|l|c|}
\hline
Parameter & $B^{\pm}\to DK^{\pm}$ \\
\hline
$x_-$ & $+0.095\pm0.045\pm0.014\pm0.010$ \\
$y_-$ & $+0.137^{+0.053}_{-0.057}\pm0.015\pm0.023$ \\
corr($x_-,y_-$) & $-0.315$ \\
\hline
$x_+$ & $-0.110\pm0.043\pm0.014\pm0.007$ \\
$y_+$ & $-0.050^{+0.052}_{-0.055}\pm0.011\pm0.017$ \\
corr($x_+,y_+$) & $+0.059$ \\
\hline
\end{tabular}
\end{center}
\end{table}

We perform a combined likelihood fit~\cite{midp} to four signal
selection variables in all DP bins for the $B^{\pm}\to DK^{\pm}$
signal and Cabibbo-favored $B^{\pm}\to D\pi^{\pm}$ control samples;
the free parameters of the fit are $x_{\pm}$, $y_{\pm}$, overall
normalization (see \eqref{midp}) and background fraction.
Table~\ref{tab:midp} summarizes the results obtained for
$B^{\pm}\to DK^{\pm}$. From these results we obtain $\phi_3
=(77.3^{+15.1}_{-14.9}\pm4.1\pm4.3)^\circ$ and $r_B=0.145\pm0.030
\pm0.010\pm0.011$, where the first error is statistical, the
second is systematic, and the last error is due to limited
precision on $c_i$ and $s_i$. Although $\phi_3$ has a mirror
solution at $\phi_3+180^\circ$, we retain the value consistent
with $0^\circ<\phi_3<180^\circ$. We report evidence for direct
$\CP$ violation, the fact that $\phi_3$ is nonzero, at the $2.7$
standard deviation ($\sigma$) level. Compared to results of the
model-dependent DP method~\cite{dalitz}, this measurement has
somewhat poorer statistical precision owing to two factors: (a)
the error itself is inversely proportional to $r_B$, the central
value of which has gotten smaller in this analysis and (b) the
binned approach is expected to have on average $10-20\%$
poorer result than the unbinned one~\cite{bonder}. On the
positive side, however, the large model uncertainty for the
model-dependent study ($8.9^\circ$) is now replaced by a purely
statistical uncertainty due to limited size of the $\psi(3770)$
data sample available at CLEO ($4.3^\circ$). The model-independent
approach therefore offers an ideal avenue for Belle II and
LHCb~\cite{lhcb} in their pursuits of $\phi_3$.

\subsection{GLW and ADS methods}

The first two methods proposed for the measurement of $\phi_3$
(Section~\ref{sec:phi3}) also go by the first initials of the
authors name: (a) ``GLW'' for Gronau-London-Wyler~\cite{glw} and
(b) ADS for Atwood-Dunietz-Soni~\cite{ads}. Although these methods
are not as competitive as the Dalitz method, they provide useful
complementarity to the final constraint on $\phi_3$. For the GLW
method, Belle has performed an analysis of $B^{\pm}\to DK^{\pm}$
using channels in which the neutral $D$ meson decays to the
$\CP$-even ($K^+K^-$ and $\pi^+\pi^-$) and $\CP$-odd [$\KS\pi^0$,
$\KS\omega(\to\pi^+\pi^-\pi^0)$, $\KS\eta(\to\gamma\gamma$ and
$\pi^+\pi^-\pi^0)$ and $\KS\eta^{\prime}(\to\eta\pi^+\pi^-$ and
$\rho^0\gamma)$] states. The physics observables are
\begin{eqnarray}
R_{\CP\pm}&=&2\,\frac{\Gamma(B^-\to D_{\CP\pm}K^-)+\Gamma(B^+\to D_{\CP\pm}K^+)}
{\Gamma(B^-\to D_{\rm fav}K^-)+\Gamma(B^+\to D_{\rm fav}K^+)},\\ \nonumber
A_{\CP\pm}&=&\frac{\Gamma(B^-\to D_{\CP\pm}K^-)-\Gamma(B^+\to D_{\CP\pm}K^+)}
{\Gamma(B^-\to D_{\CP\pm}K^-)+\Gamma(B^+\to D_{\CP\pm}K^+)},
\label{eq:glw}
\end{eqnarray}
where $D_{\rm fav}$ denotes the Cabibbo-favored decay mode for the $D$
meson such as $D^0\to K^-\pi^+$. These four observables are functions
of $\phi_3$, $r_B$ and $\delta_B$. Using the full $\Y4S$ data sample,
we obtain $R_{\CP+}=1.03\pm0.07\pm0.03$, $R_{\CP-}=1.13\pm0.09\pm0.05$,
$A_{\CP+}=+0.29\pm0.06\pm0.02$, and $A_{\CP-}=-0.12\pm0.06\pm0.01$, where
the quoted errors are statistical and systematic, respectively. There is
clear evidence of direct $\CP$ violation ($A_{\CP}\neq 0$) for the
$\CP$-even modes. In the case of ADS method our study extends to
$B^-\to D^*K^-$, $D^*\to D\gamma\, (D\pi^0)$ where the $D$ meson is
reconstructed in the doubly Cabibbo suppressed channel $D\to K^+\pi^-$.
We report the first evidence for signal in this mode with a $3.5\sigma$
significance.

\section{Study of {\boldmath $B\to\eta h\,(h=K,\pi)$} decays}

\begin{figure}
\begin{center}
\includegraphics[width=0.6\columnwidth]{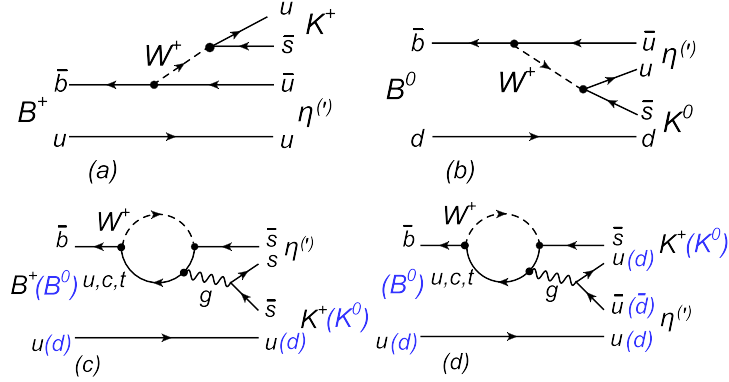}
\end{center}
\caption{(a) $b\to u$ tree diagram for $B^+\to\eta^{(\prime)}K^+$, (b)
color-suppressed $b\to u$ tree diagram for $B^0\to\eta^{(\prime)}K^0$,
and (c),(d) $b\to s$ penguin diagrams for $B\to\eta^{(\prime)}K$.}
\label{fig:etah1}
\end{figure}

The $B^{\pm}\to\eta^{(\prime)}h^{\pm}\,(h=K,\pi)$ decays proceed via $b
\to s$ penguin processes and Cabibbo-suppressed $b\to u$ tree transition,
as shown in Fig.~\ref{fig:etah1}. The large $B\to\eta^\prime K$ and small
$B\to\eta K$ branching fractions can be thought of as an artifact of
$\eta-\eta^\prime$ mixing together with the constructive and destructive
interference between the two penguin diagrams. On the other hand, owing to
small contribution from the color-suppressed tree in case of $B^0\to\eta
K^0$, its branching fraction is expected to be lower than that of $B^+\to
\eta K^+$. Concerning direct $\CP$ violation, one could find sizeable
effects in $B^{\pm}\to\eta h^{\pm}$, proportional to the degree of
interference between the contributing penguin and tree amplitudes.

Using the entire $\Y4S$ data sample of Belle, we have studied $B\to\eta
h$ decays~\cite{etah} where the $\eta$ meson is reconstructed in the
two channels $\eta\to\gamma\gamma$ and $\eta\to\pi^+\pi^-\pi^0$. The
measured branching fractions are $\BF(B^{\pm}\to\eta K^{\pm})=[2.12\pm
0.23\stat\pm0.11\syst]\times10^{-6}$, $\BF(B^{\pm}\to\eta\pi^{\pm})=
[4.07\pm0.26\stat\pm0.21\syst]\times10^{-6}$ and $\BF(B^0\to\eta K^0)=
[1.27^{+0.33}_{-0.29}\stat\pm0.08\syst]\times10^{-6}$. The dominant
systematic errors are due to uncertainties on reconstruction efficiencies
of the $\eta$, $\pi^0$ and $\KS$ mesons. The $B^0\to\eta K^0$ decay
is observed for the first time with a significance of $5.4\sigma$.
We also find evidence for $\CP$ violation in the charged $B$ decay
channels, $A_{\CP}(B^{\pm}\to\eta K^{\pm})=-0.38\pm0.11\stat\pm0.01\syst$
and $A_{\CP}(B^{\pm}\to\eta\pi^{\pm})=-0.19\pm0.06\stat\pm0.01\syst$
with significances of $3.8\sigma$ and $3.0\sigma$, respectively. All
these branching fraction and $\CP$ measurements supersede our earlier
results~\cite{oldetah}.

\section{Conclusions and future prospect}

Among the sample of results we presented at this prestigious conference
of Moriond include the most precise determination of $\sin(2\phi_1)$
in the $B\to (c\cbar)K^0$ decays, a novel method to determine the same
parameter in the $\Y5S$ data, a first model-independent DP analysis
to determine the UT angle $\phi_3$, and evidence for direct $\CP$
violation in $B\to\eta h$. The second and third measurements are
more of a proof-of-principle in nature, and hold a great promise for
experiments at the super flavor factories. At the moment, there are
many ongoing $\CP$ related analyzes, such as the UT angle $\phi_2$,
with the full $\Y4S$ statistics of Belle. Therefore we expect many
interesting results to come out soon while one is waiting for the next
phase of the experiment, Belle II.

\section*{Acknowledgments}
We thank Y. Kwon, M. Nakao, Y. Sato, Y. Sakai and K. Trabelsi for
their helps during the preparation of the talk and proceedings.

\end{document}